# Frequency Response Study on the ERCOT under High Photovoltaic (PV) Penetration Conditions

Xuemeng Zhang, Shutang You, Yong Liu
Dept. of Electrical Engineering and Computer Science
The University of Tennessee, Knoxville
Knoxville, TN, USA
syou3@utk.edu

Yilu Liu
The University of Tennessee, Knoxville, and Oak Ridge National Lab
TN, USA
liu@utk.edu

*Abstract*— Solar photovoltaic (PV) generation is growing rapidly around the world. However, PV generation, based on inverter, is fundamentally different from conventional synchronous generators. It is of vital importance to understand the impacts of increased penetration of PV generation on power system dynamic performance. This paper investigates frequency response of the Electric Reliability Council of Texas (ERCOT) with high PV penetration in the future year. To start with, a realistic baseline dynamic model is validated using synchrophasor measurements. A dynamic simulation is performed to evaluate the impacts of high PV generation on frequency response.

*Index Terms*-- Solar photovoltaic (PV); Frequency response; Power system model validation; Wide-area measurement

## I. Introduction

The recent years have seen substantial growth in photovoltaic (PV) generation. According to a study [1], PV generation has the potential to generate up to 14% of the nation's total electricity demand by 2030 and 27% by 2050. Meanwhile, the U.S. Energy Information Administration projects that a total of 60-GW capacity will retire by 2020 [2]. The displacement of the conventional synchronous generators is undoubtedly cutting back the national carbon footprint, but simultaneously results in numerous challenges in power system operation and planning. One of the major concerns is primary frequency response degradation. Solar plants, by their design, do not provide primary frequency control. Additionally, they do not have inertia compared to conventional synchronous generators. Thus, systems with reduced total inertia experience a sharper immediate frequency drop during under-frequency events and are more vulnerable to involuntary under-frequency load shedding. Given these concerns, this paper aims to evaluate the primary frequency response of the ERCOT system under different PV penetration conditions.

The frequency response studies have also been conducted on the large-scale interconnected power grids to evaluate the impacts of high renewable penetration. [3] investigated frequency response with high penetration of wind and PV in South Australia. It is concluded that low inertia and secondary PV tripping can become serious issues for network frequency regulation and in some situations they can even cause system oscillation and stability concerns. The authors in [4] investigated the primary frequency response adequacy of the U.S. Eastern Interconnection with high-wind penetration in the year 2030. In one WECC study, the existing thermal units are displaced by from 15% to 18% wind penetration and the simulation results indicate that the combination of wind inertia control and governor control can improve system nadir and settling frequency [5]. Simulations on an IEEE 39 bus system were carried out in [6] to address the various potential impacts of high level penetration PV. Observations were made regarding the changes in unit commitment strategies as well as dispatch flexibility.

This paper is organized as follows. Section II describes the basic information about studied system and how to build various PV penetration level scenarios. In section III, the frequency response of different PV penetration levels is investigated. Conclusions derived from these analyses are presented in Section IV.

## II. Scenario Development

As mentioned, Electric Reliability Council of Texas (ERCOT) will be used in this paper for case studies. In this section, this bulk power grid and its models and simulation scenarios will be briefly introduced.

### A. Model Overview

ERCOT system is one of the three major electrical grids in the North America. It covers most of the state of Texas and has about 74 GW for peak demand [7]. In this paper, the 2015 summer peak model from Electric Reliability Council of Texas (ERCOT) is used. The basic information of this system is given in Table I.

TABLE I. BASIC INFORMATION OF ERCOT SYSTEM

| Total Load | MW | 74127.61 |
|---|---|---|
|  | MVAr | 18636.42 |
| Total Generation | MW | 75735.83 |
|  | MVAr | 19845.34 |
| Total Number of Generators | | 690 |
| Total Number of Buses | | 6240 |
| Total Number of Branches | | 6320 |

This work made use of Engineering Research Center Shared Facilities supported by the Engineering Research Center Program of the National Science Foundation and the U.S. Department of Energy under NSF Award Number EEC-1041877 and the CURENT Industry Partnership Program.

## B. Model Validation Using Synchrophasor Frequency Measurement

Wide area measurement system (WAMS) has been deployed in ERCOT and it provides accurate measurements of both systems' frequency performances. With the help of WAMS, it has been noticed that the original ERCOT models tend to exhibit frequency responses that are more optimistic than actual performances measured by WAMS [8]. Therefore, as the first step of this study, the ERCOT models are validated using WAMS frequency measurement to further enhance the model credibility.

Consider the frequency response metrics, rate of change of frequency is determined by the amount of rotating mass (mechanical inertia) in the interconnection; the combination of inertia and Primary Frequency Response dictates frequency nadir; after the frequency decline has been arrested, continued delivery of Primary Frequency Response will stabilize frequency at a steady-state settling level. Therefore, the proper modeling of mechanical inertia and governor has significant impact on the accurate representation of the ERCOT system frequency response.

Dominant governor type in the original ERCOT models is steam turbine-governor (TGOV1). In its block diagram (Fig. 1), the ratio, $T_2/T_3$, equals the fraction of turbine power that is developed by the high-pressure turbine. $T_3$ is the reheater time constant, and $T_1$ is the governor time constant. Usually, the range of parameter $T_1$ is very small so that $T_1$ is seldom changed. Whereas, $T_2$ and $T_3$ can be easily modified in a certain range.

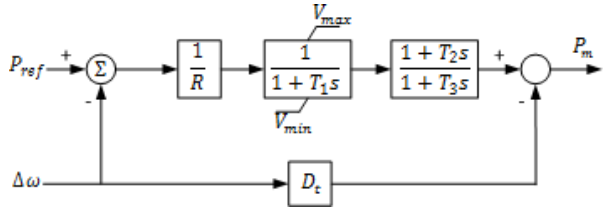

Figure 1. TGOV1 block diagram

As demonstrated above, the governor capacity, mechanical inertia and reheater time constant could account for the frequency response mismatch between simulation and measurement in the ERCOT system. To validate the dynamic models, a three-step approach of adjusting these key factors is applied:

1) Reducing governor capacity
2) Modifying reheater time constant
3) Calibrating mechanical inertia

## C. Case Study

Five actual events were studied to prove the accuracy of the validated dynamic models (Table III). An example is demonstrated as following, which occurs at 16:30:20 Coordinated Universal Time (UTC) on January 8, 2016. Measurements from one FDR and validated model results are plotted in Fig. 2.

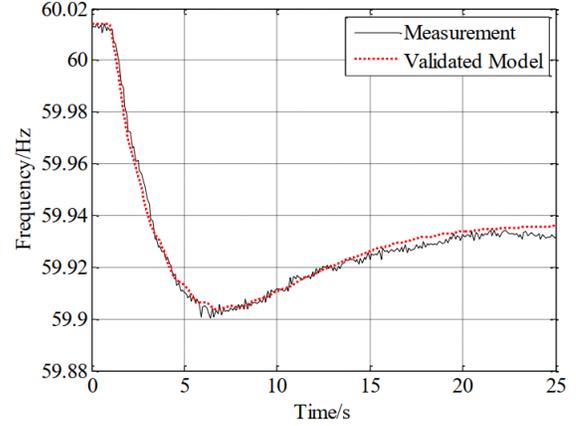

Figure 2. ERCOT model frequency response validation results

TABLE II. COMPARISON OF MEASUREMENT AND SIMULATION

|  | FNET/Grid Measurement | Validated Model Simulation | Difference |
|---|---|---|---|
| Frequency nadir (Hz) | 59.902 | 59.903 | 0.001 |
| Rate of change of frequency (mHz/s) | 23 | 30 | 7 |
| Frequency settling time (s) | 21 | 23 | 2 |
| Settling frequency (Hz) | 59.932 | 59.935 | 0.003 |

As can be observed in Fig. 2, by reducing governor capacity, justifying governor re-heater time constant and mechanical inertia, the rate of change of frequency and frequency nadir and the settling frequency could be matched well. Table II shows difference between measurement and simulation results.

Besides, average difference between measurement and simulation for all five events are shown in Table IV. Results indicate that simulated frequency response of all these five events listed in Table III match well with the actual measurement, which further prove the accuracy of the validated dynamic models.

TABLE III. INFORMATION OF ACTUAL EVENT

| FNET/Grid Measurement | Time | Generation Trip (MW) |
|---|---|---|
| 1 | 2015/12/03 05:25:44 | 360 |
| 2 | 2015/12/21 21:19:53 | 320 |
| 3 | 2015/12/30 17:17:18 | 540 |
| 4 | 2016/01/08 16:30:20 | 390 |
| 5 | 2016/01/22 05:14:43 | 660 |

TABLE IV. AVERAGE DIFFERENCE OF MEASUREMENT AND SIMULARION IN FIVE CASES

| Cases | Frequency nadir (Hz) | Rate of change of frequency (mHz/s) | Frequency settling time (s) | Settling frequency (Hz) |
|---|---|---|---|---|
| 1 | 0.01 | 6 | 3 | 0.006 |
| 2 | 0.01 | 1 | 1 | 0.007 |
| 3 | 0.003 | 2 | 0 | 0.004 |
| 4 | 0.003 | 7 | 2 | 0.003 |
| 5 | 0 | 5 | 1 | 0.009 |
| Average | 0.005 | 4.2 | 1.4 | 0.006 |

D. *Scenario Development*

Define abbreviations and acronyms the first time they are used in the text, even after they have been defined in the abstract. Abbreviations such as IEEE, SI, ac, dc, and rms do not have to be defined. Do not use abbreviations in the title or section headings unless they are unavoidable.

For wind power is the most significant renewable generation sources in ERCOT, it would be simply not realistic if only solar power is simulated for frequency response studies. Therefore, both wind and solar power should be modeled in this paper's simulation scenarios. Furthermore, for a bulk power grid such as ERCOT, solar power location matters. Therefore, it is important to consider the realistic solar power distribution in the case studies. In this study, the wind and solar PV distribution information from the U.S. Department of Energy Wind Vision Study was used to develop three scenarios for ERCOT. The total renewable (including wind and solar PV) penetration rate is chosen to be 20%, 40%, and 60%. These three levels of renewable penetration should be able to largely reflect ERCOT renewable integration in the following several decades. As shown in Table V, different PV penetration levels with fixed wind penetration are defined. Wind power is modeled as doubly-fed electric machine (DFIG)-based wind farms in this study since it is the most common wind farm type right now.

TABLE V. WIND AND SOLAR PV PENETRATION RATES IN ERCOT

|  | Wind | PV |
|---|---|---|
| 20% | 15% | 5% |
| 40% | 15% | 25% |
| 60% | 15% | 45% |

To build high renewable penetration scenarios, a portion of conventional generators are replaced with PV and wind plants. According to ERCOT existing wind farm, active wind capacity reaches 10 GW or so, most of which locate in West Texas. And PV plants would be evenly distributed throughout the ERCOT. Fig. 3-5 show the PV and wind plants locations of 20%, 40% and 60% penetration levels. In 20% scenario, most PV locate in West and South, which is consistent with projected PV distribution in [9]. In 40%, 60% scenarios, PV are much more evenly distributed.

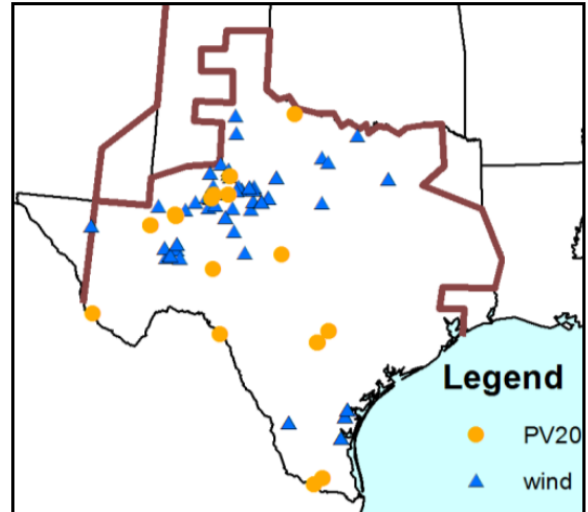

Figure 3. 20% renewable distribution map

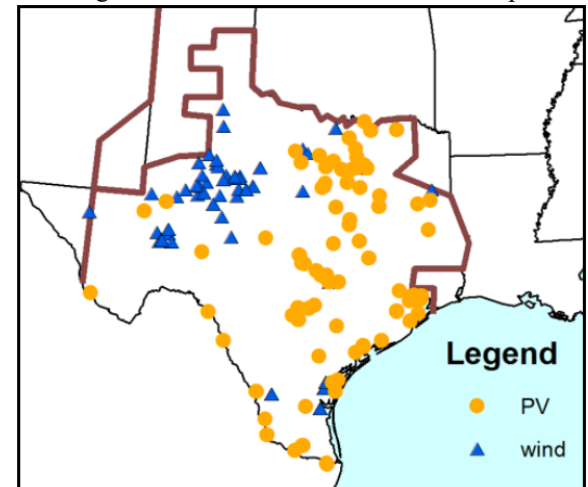

Figure 4. 40% renewable distribution map

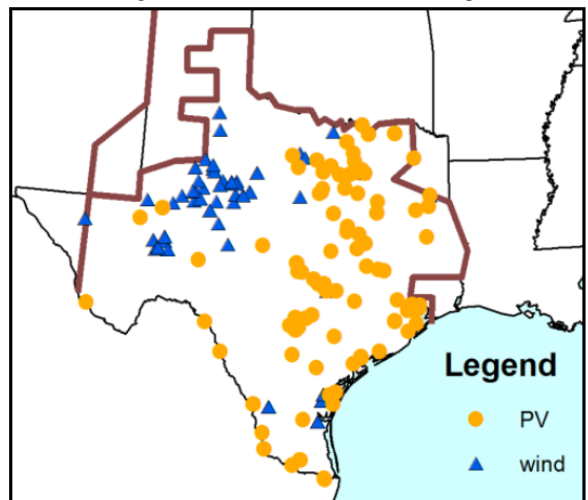

Figure 5. 40% renewable distribution map

III. CASE STUDY OF IMPACT ON FREQUENCY RESPONSE

In this section, using the scenarios developed in Section II, the impacts of renewable power on the ERCOT frequency response are studied. The frequency responses of the ERCOT system after a generation disturbance of 1129 MW is given in

Fig. 6 and the frequency metrics are given in Table VI. As shown in Fig. 6 and Table VI, the frequency response deteriorates gradually due to the renewable integration. For example, in the 60% case, the frequency nadir has been reduced to 59.26 Hz, which is already lower than the first stage ERCOT UFLS frequency threshold (59.3 Hz) [10]. This means ERCOT has to take countermeasure in order to incorporate 60% renewable, otherwise, more UFLSs in the ERCOT system will be seen. This observation also indicates that it is more challenging for small power systems to deal with the high renewable impact on frequency responses.

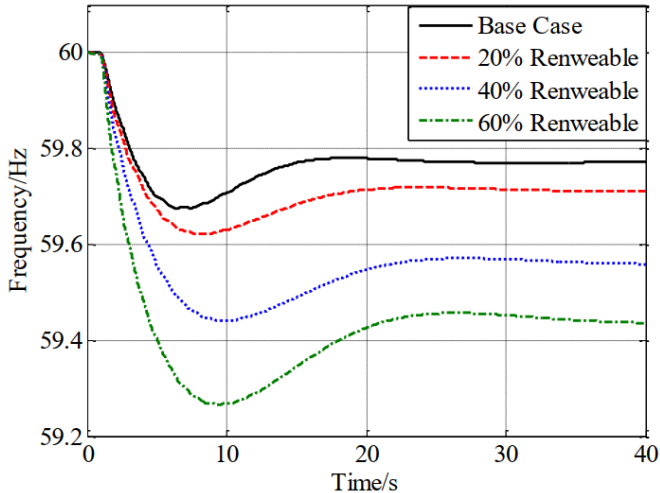

Figure 6. ERCOT frequency response change due to renewable integration

TABLE VI. ERCOT FREQUENCY RESPONSE METRICS CHANGE DUE TO RENEWABLE GENERATION

| Frequency response metrics | Base case | 20% | 40% | 60% |
|---|---|---|---|---|
| ROCOF (mHz/s) | 100 | 110 | 140 | 200 |
| Nadir (Hz) | 59.67 | 59.62 | 59.44 | 59.26 |
| Settling time (s) | 15 | 20 | 23 | 25 |
| Settling frequency (Hz) | 59.77 | 59.71 | 59.55 | 59.43 |

## IV. CONCLUSION

In this paper, the impact of PV generation on frequency response is studied for the ERCOT system. By adjusting active governor capacity and mechanical inertia, a realistic baseline dynamic model is validated using synchrophasor measurements. A dynamic simulation is performed to evaluate the impact of high PV generation on frequency response.

ACKNOWLEDGMENTS

This work made use of Engineering Research Center shared facilities supported by the Engineering Research Center Program of the National Science Foundation and the Department of Energy under NSF Award Number EEC-1041877 and the CURENT Industry Partnership Program.